\documentclass{ws-procs9x6}

\newcommand{\lsim}{\lower0.6ex\vbox{\hbox{$ \buildrel{\textstyle <}\over{\sim}\ $}}}
\newcommand{\gsim}{\lower0.6ex\vbox{\hbox{$ \buildrel{\textstyle >}\over{\sim}\ $}}}

\newcommand{\hkpc}{h^{-1}\mathrm{kpc}}

\newcommand{\hMsun}{\ h^{-1}\mathrm{M}_{\odot}}

\newcommand{\hMpck}{\ h\mathrm{Mpc}^{-1}}

\newcommand{\mpt}{m_{\mathrm{p}}}
\newcommand{\GeV}{\mathrm{GeV}}

\newcommand{\Omegam}{\Omega_{M}}

\newcommand{\sig}{\sigma_{8}}

\newcommand{\Rvir}{R_{\mathrm{vir}}}

\newcommand{\Vmax}{V_{\mathrm{max}}}

\newcommand{\Mchi}{M_\chi}

\newcommand{\beq}{\begin{equation}}
\newcommand{\eeq}{\end{equation}}

\newcommand{\dd}{\mathrm{d}}

\begin{document}

\title{Dark Matter Halos: Shapes, The Substructure Crisis, 
and Indirect Detection}

\author{
A.~R. Zentner$^1$, 
S.~M. Koushiappas$^2$, and 
S. Kazantzidis$^{1\mathrm{,}3}$}

\address{
$^{1}$Kavli Institute for Cosmological Physics \& 
Department of Astronomy and Astrophysics, The University of Chicago, 
Chicago, IL 60637 USA \\
$^{2}$Department of Physics, Swiss Federal Institute of Technology, 
ETH H{\"{o}}nggerberg, Z{\"{u}}rich, Switzerland \\
$^{3}$Institute for Theoretical Physics, 
University of Z{\"{u}}rich, Z{\"{u}}rich, Switzerland
}

\maketitle

\abstracts{
In this proceeding, we review three recent results.  
First, we show that halos formed in simulations with gas 
cooling are significantly rounder than halos formed in 
dissipationless $N$-body simulations.  The increase in 
principle axis ratios is $\sim 0.2 - 0.4$ in the inner 
halo and remains significant at large radii.
Second, we discuss the CDM substructure crisis and 
demonstrate the sensitivity of the crisis to the spectrum of 
primordial density fluctuations on small scales.  Third, 
we assess the ability of experiments like VERITAS and GLAST to 
detect $\gamma$-rays from neutralino dark matter annihilation 
in dark subhalos about the MW.
}

%%%%%%%%%%%%%%%%%%%%%%%%%%%%%%%%%%%%%%%%%%%%%%%%%%%%%%%%%%%%%%%%%%%%%%%
\section{Introduction}

A proponderance of evidence indicates that galaxies 
are embedded in massive, extended dark matter (DM) {\em halos}.  
Simulations of structure formation in the hierarchical 
cold dark matter (CDM) paradigm predict that CDM halos 
are generally triaxial\cite{triaxial,dubinski_carlberg91} 
that they teem with self-bound {\em subhalos}\cite{dsp}.  

The structure of halos is an important ingredient 
in modeling the DM direct detection signals\cite{ddm} 
and halo shapes have recently received attention for 
testing the CDM paradigm as new and 
improved probes of halo shape have 
been applied\cite{hs,sag}.  
{\em Dissipationless} simulations predict that 
Milky Way(MW)-size halos have a mean minor-to-major 
axis ratio of $c/a \approx 0.6-0.7$ with a 
dispersion of $\sim 0.1$\cite{triaxial}, while dynamical
studies suggest that the observed coherence of the 
Sagittarius tidal stream constrains the inner MW 
halo to $c/a \gsim 0.8$\cite{sag}. 
In \S~\ref{sec:shapes}, we present recent 
results on the effect of baryonic dissipation on 
halo shapes in high-resolution, cosmological simulations.

In \S~\ref{sec:subs}, we turn to halo substructure.  
In the MW and M31, there are more than an order of 
magnitude fewer observed satellites than the 
predicted number of subhalos of comparable size\cite{dsp}.  
Several explanations have been offered, including 
alternative DM properties\cite{sdm} and 
inefficient galaxy formation in the shallow 
potentials of small subhalos\cite{bkwl}.  
We study the sensitivity of the dwarf satellite 
population to the primordial power spectum 
(PPS) of density fluctuations on small, sub-galactic 
scales and demonstrate that our interpretation 
of the missing satellite problem is a function 
of the amount of small-scale power.  
If the lack of luminous MW satellites is 
due to inefficient galaxy formation, the 
MW halo should contain $\gsim 10^{2}$ otherwise 
dark subhalos.  Strong lensing will be one probe of 
dark subhalos\cite{dk}.  More speculatively, 
the annihilation of DM particles in these 
dense substructures may result in numerous $\gamma$-ray 
sources in the MW halo.  We assess the potential 
for instruments like VERITAS\cite{veritas} and 
GLAST\cite{glast} to detect such sources in favorable 
models of supersymmetric (SUSY) DM in \S~\ref{sec:idds}.

\section{Halo Shapes}
\label{sec:shapes}

%%%%%%%%%%%%%%%%%%%%%%%%%%%%%%%%%%%%%%%%%%%%%%%%%%%%%%%%%%%%%%%
\begin{figure}[t]
\centerline{\epsfxsize=5.0in\epsfbox{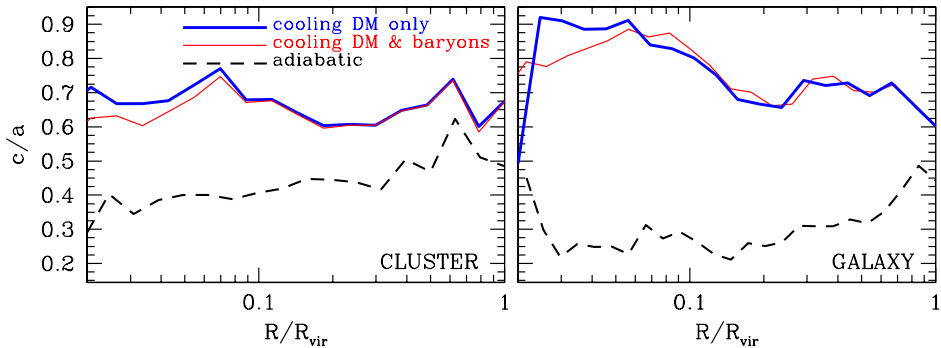}}
\label{fig:schp}
\caption{
The effect of gas cooling on halo shapes. 
{\em Left}:  Minor-to-major axis ratio $c/a$, as a 
function of major axis length for a cluster-size halo.  
The {\em dashed} line shows the shape profile of the DM 
in the adiabatic simulation.  The {\em thick, solid} line shows 
the shape profile of DM in the cooling 
run, while the {\em thin, solid} line shows the shape 
profile for DM and baryons in the cooling run.  
{\em Right}: Same as the left panel, but for a  
MW-size galaxy progenitor (see text).
}
\end{figure}
%%%%%%%%%%%%%%%%%%%%%%%%%%%%%%%%%%%%%%%%%%%%%%%%%%%%%%%%%%%%%%%

We studied the effect of gas cooling on the shapes of 
DM halos using high-resolution cosmological 
simulations of cluster and galaxy formation 
in a concordance $\Lambda$CDM cosmology.  
The simulations were performed with the ART 
$N$-body plus Eulerian gasdynamics code\cite{art}.
We refer the reader to Kazantzidis et al.\cite{gas} for further details.

Briefly, we analyzed simulations of $8$ cluster-size objects 
of mass $10^{13} \hMsun$ to $3 \times 10^{14} \hMsun$.  
The cluster simulations had a peak force resolution 
of $\simeq 2.4 \hkpc$ and a DM particle mass of 
$\mpt \simeq 2.7 \times 10^{8} \hMsun$.  
We also analyzed a simulation of the early evolution 
($z \gsim 4$) of a galaxy that becomes 
MW-size at $z=0$ described by Kravtsov\cite{kravtsov03}.  
This simulation had $\mpt \simeq 9.2 \times 10^{5} \hMsun$ 
and peak resolution $\simeq 183 \hkpc$.  The mass and 
force resolution are adequate to study the inner regions 
of halos reliably.  For each object,  
we analyzed two sets of simulations started
from the same set of initial conditions, 
but including different physical
processes.  In one set, the gas dynamics were treated 
adiabatically, without any radiative cooling and the results agreed 
well with those of $N$-body simulations with no baryonic component. 
The second set of simulations included radiative cooling, 
and star formation. 

We measured halo shapes by diagonalizing the moment of inertia 
tensor\cite{dubinski_carlberg91}.  
We used ``differential'' shape measurements 
because this makes the axis ratios measured at each radial 
bin nearly independent.  Our main results are summarized in 
Figure~1.  In the left panel, we show the profile $c/a$, 
as a function of major axis length for a representative 
cluster-size halo. On the right, we show results for the galaxy 
progenitor.  The net effect of baryon dissipation is striking.  
At small radii, the axis ratios in the cooling 
simulations are greater by $\Delta(c/a) \gsim 0.3$ 
and the systematic difference persists out to 
$\sim \Rvir$, where $\Delta(c/a) \sim 0.1$.  
The baryons in the cluster are mostly in 
a massive, central, elliptical galaxy while in the 
galaxy formation simulation, 
$\sim 90\%$ of the baryons are in a flattened, 
gaseous disk. In both cases the effect of cooling 
is weakly dependent upon radius implying that 
the effect of baryonic dissipation on halo shapes is not 
critically sensitive to the detailed morphology of the 
baryonic component.  In addition, the axis ratios change 
with radius in a manner that is not generally monotonic, 
indicating that different regions of a system may be 
flattened to different degrees.

%%%%%%%%%%%%%%%%%%%%%%%%%%%%%%%%%%%%%%%%%%%%%%%%%%%%%%%%%%%%%%%%%
\section{Halo Substructure}
\label{sec:subs}

The most accurate technique for studying halo 
substructure is numerical simulation; 
however, the computational expense of simulations 
limits their dynamic range and their applicability 
in explorations of cosmological parameter space.  
To overcome this, Zentner and Bullock (ZB)\cite{zb03} 
developed an approximate, analytic model for subhalo 
populations and an updated model has recently been 
successfully tested against a suite of 
$N$-body simulations\cite{z05}.  The model 
approximately accounts for the merger statistics 
of subhalos, dynamical friction, and 
mass loss and redistribution due to tidal forces.  
The model allows one to generate hundreds of 
realizations of MW-like halos and thereby explore 
the distribution of possible subhalo populations.  

%%%%%%%%%%%%%%%%%%%%%%%%%%%%%%%%%%%%%%%%%%%%%%%%%%%%%%%%%%%%%%%%%
\begin{figure}[t]
\centerline{\epsfxsize=3.5in\epsfbox{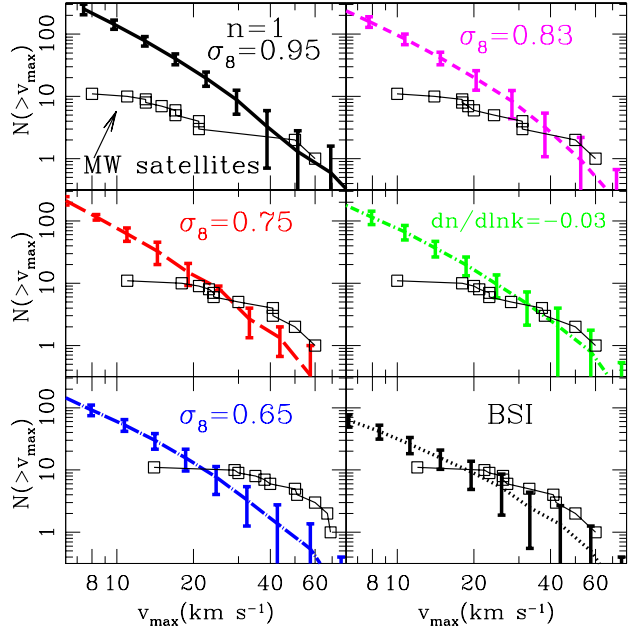}}
\label{fig:dsp}
\caption{
Dwarf satellites and the power spectrum.  
We show the observed satellite velocity 
functions ({\em squares}) and the predicted 
satellite velocity functions ({\em thick lines}) 
for $6$ different PS.  Clockwise from the top 
left: 
standard $n=1$, $\sig = 0.95$; 
$n=0.94$, $\sig = 0.83$; 
WMAP best-fit $n=1.03$, $\dd n/\dd \ln k = -0.03$, $\sig = 0.84$; 
BSI; 
$n=0.84$, $\sig = 0.65$; and 
$n=0.90$, $\sig = 0.75$.  The models are labeled by 
$\sig$. Lines are the means of $100$ model 
realizations and errorbars represent the $1\sigma$ scatter.
Observational data are from the review of Mateo$^{21}$.
}
\end{figure}
%%%%%%%%%%%%%%%%%%%%%%%%%%%%%%%%%%%%%%%%%%%%%%%%%%%%%%%%%%%%%%%%%%

In the standard paradigm, structure forms from 
primordial density fluctuations characterized by 
a nearly scale-invariant PPS, $P(k) \propto k^{n}$ 
with $n \simeq 1$.  This basic picture has 
significant observational support\cite{wmap}.  
However, cosmic microwave background anisotropy 
constrains the PPS on large scales, 
$k \sim 10^{-2} \hMpck$, while halo substructure 
is sensitive to small scale power, $k \sim 10-100 \hMpck$.  
ZB studied the effect of variant 
power spectra on the MW dwarf satellites.  They 
took several PPS with various motivations, all 
normalized to COBE: (1) standard $n=1$, $\sig = 0.95$; 
(2) $n=0.94$, $\sig = 0.83$; 
(3) $n=0.9$, $\sig = 0.75$; 
(4) running mass inflation $n=0.84$, $\sig = 0.65$; 
(5) broken scale-invariance (BSI) with a power cut-off at 
$k_c = 1 \hMpck$\cite{kl}; and 
(6) the best-fit running spectrum from WMAP 
$n=1.03$, $\dd n/\dd \ln k = -0.03$, $\sig = 0.84$.
The steps in the calculation are first to generate 
MW halo substructure realizations for each PPS 
and to model the velocity dispersions of the embedded 
stellar components to determine the appropriate 
subhalo size (labelled by maximum circular velocity 
$\Vmax$) in which the observed satellites may be 
embedded.  In this way, one constructs predicted 
and observed cumulative velocity functions.

Figure~2 summarizes the results.  
First, one sees that the degree to 
which the dwarf satellite problem represents a 
challenge is greatly alleviated in the WMAP best-fit 
cosmology.  The level at which inefficient 
galaxy formation or a critical mass scale for 
galaxy formation must be invoked to solve the 
satellite scarcity problem is degenerate with 
the PPS on small scales.  Second, the 
MW satellite population by itself provides 
independent evidence against extreme models, 
such as the low normalization, $\sig = 0.65$ 
model which under-predict substructure.

%%%%%%%%%%%%%%%%%%%%%%%%%%%%%%%%%%%%%%%%%%%%%%%%%%%%%%%%%%%%%%%%%%
\section{$\gamma$-rays from Dark Substructure}
\label{sec:idds}

One way of probing the distribution and properties 
of substructure as well as the particle nature 
of the DM is through the detection of 
gamma-rays from annihilations  
of the dark matter particle 
in the dense, inner regions of subhalos. 
The currently favored DM particle is 
provided by supersymmetry (SUSY) and it is 
the lightest of the neutralinos ($\chi$). 
The uncertainties involved in trying to deduce information 
about the distribution and properties of substructure 
indirectly via the detection of $\gamma$-rays are twofold.  
First, there are uncertainties that stem from the underlying 
cosmological model and the details of formation of very 
small-scale structures\cite{zb02,zb03} and second, 
uncertainties that arise from the lack of knowledge of the 
mass and couplings of the dark matter particle. 
Using the analytic substructure 
model of \S~\ref{sec:subs}, we can assess 
the ability of experiments like VERITAS 
and GLAST to detect $\gamma$-ray fluxes from 
DM annihilations.  

Koushiappas et al.\cite{kzb04}, adopted this approach and 
assumed the most optimistic SUSY parameters consistent
with constraints on $\Omegam$\cite{wmap} to determine  
the number of expected detections at a significance 
$S > 3$, as a function of subhalo mass $M$.  In 
order to project counts of observed subhalos beyond 
the masses of the dwarf galaxies, several 
physically-motivated extrapolations are necessary; however, 
the recent simulation of ``mini-halos'' at $z \sim 26$ are 
a first step toward justifying these extrapolations with 
explicit numerical simulations\cite{mini}.

Our results are summarized in Figure~3.  The figure shows 
that for $\chi$ masses $\Mchi \lsim 100 \GeV$, the large 
field of view of GLAST and the energy sensitivity of 
VERITAS will allow them to detect substructure when operated 
in concert. For example, if $\Mchi \sim 75 \GeV$, then 
in the case of optimal coupling to photons there will be 
on average $\sim 1$ detectable subhalo per GLAST field of view. 
In this case, subsequent direct observations with VERITAS 
should be able to confirm the line emission feature 
at an energy of $\sim \Mchi$ 
after an exposure time of $\sim 450$ hours. 
For $100 \GeV \lsim \Mchi \lsim 500 \GeV$, detection requires 
an instrument with a large effective area, like VERITAS; however, 
such a detection must rely on serendipity due to the small 
number of potentially detectable objects in VERITAS' comparably 
small field of view.  For neutralino masses in excess of 
$\Mchi \gsim 500 \GeV$, substructure detection 
via the $\gamma$-ray signal is unlikely with either 
GLAST or VERITAS.\cite{kzb04}

%%%%%%%%%%%%%%%%%%%%%%%%%%%%%%%%%%%%%%%%%%%%%%%%%%%%%%%%%%%%%%%%%
\begin{figure}[t]
\centerline{\epsfxsize=5.in\epsfbox{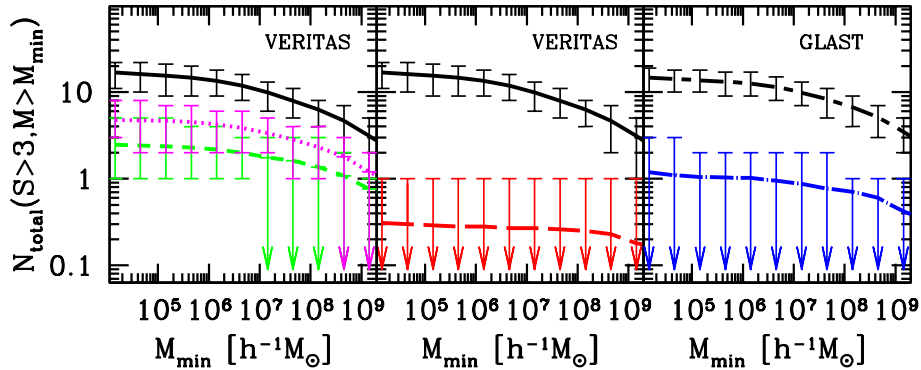}}
\label{fig:gray}
\caption{
The cumulative number of subhalos of mass $M \ge M_{\min}$ 
detectable at $S > 3$ on the sky.  
Results are based on $100$ realizations of a 
MW-size halo.  Errorbars indicate the 
$68 \%$ range and down arrows indicate that 
$> 16 \%$ realizations have zero subhalos at that mass.  
{\em Left}:  The number detectable by VERITAS.  
The {\em solid} line shows the highest detection efficiency 
case of $M_{\chi}=500 \GeV$.  For comparison, the {\em dotted} 
line shows results for $M_{\chi} = 200 \GeV$ and 
the {\em dashed} line for $M_{\chi} = 5 \mathrm{TeV}$.  
{\em Middle}:  The {\em solid} line shows our standard result for 
a $\Lambda$CDM cosmology with $n=1$ and $\sig = 0.95$.  
The {\em dashed} line shows the detectable number of 
subhalos with the WMAP best-fit running power spectrum, 
$\dd n /\dd \ln k = -0.03$.  
{\em Right}: The number detectable with GLAST.  
The {\em dashed} line represents the best case of 
$M_{\chi} = 50 \GeV$ in a standard $\Lambda$CDM cosmology.  
The {\em dot-dashed} line shows the potential number of 
detections for a $M_{\chi} = 100 \GeV$ neutralino.  
}
\end{figure}
%%%%%%%%%%%%%%%%%%%%%%%%%%%%%%%%%%%%%%%%%%%%%%%%%%%%%%%%%%%%%%%%%

\section*{Acknowledgments}

These results are based on several collaborative works.  
We thank B.~A. Allgood, J.~S. Bullock, A.~V. Kravtsov, 
B. Moore, D. Nagai, and T.~P. Walker for 
their invaluable contributions and for allowing us to 
present our results here.  We thank Von Freeman and 
Risa Wechsler for stimulating discussions.  
ARZ and SK are funded by the Kavli 
Institute for Cosmological Physics at 
The University of Chicago and 
The National Science Foundation through 
grant NSF PHY 0114422.  SMK is funded by the 
Swiss National Science Foundation.

\end{document}